\documentclass[graybox]{svmult}

\usepackage{mathptmx}       
\usepackage{helvet}         
\usepackage{courier}        
\usepackage{type1cm}        
\usepackage[psamsfonts]{amssymb}
\usepackage{amsmath}
\usepackage{amsfonts}
\usepackage{bm}

\usepackage{makeidx}         
\usepackage[dvipdfm]{graphicx}        
\usepackage{multicol}        
\usepackage[bottom]{footmisc}
\usepackage{bm}

\makeindex             

\begin{document}

\title*{Statistical mechanics of human resource allocation: 
A mathematical modeling of job-matching in labor markets}
\titlerunning{Statistical mechanics of human resource allocation}
\author{Jun-ichi Inoue and He Chen}
\institute{Jun-ichi Inoue \at Graduate School of Information Science and Technology,
Hokkaido University, N14-W-9, Kita-ku, Sapporo 060-0814, Japan \email{jinoue@cb4.so-net.ne.jp} 
\and He Chen \at Graduate School of Information Science and Technology,
Hokkaido University, N14-W-9, Kita-ku, Sapporo 060-0814, Japan \email{chen@complex.ist.hokudai.ac.jp}}
%
%
\maketitle

\abstract{
We provide a mathematical model to investigate the human 
resource allocation problem for agents, say, university graduates who are looking for their positions in labor markets. 
The basic model is described by the so-called Potts spin glass which is well-known 
in the research field of statistical physics. In the model, each Potts spin 
(a tiny magnet in atomic scale length) represents the action of each 
student, and it takes a discrete variable corresponding to the company he/she applies for. We construct the energy to include three distinct effects 
on the students' behavior, namely, collective effect, market history and international ranking of companies. In this model system, the correlations (the adjacent matrix) 
between students are taken into account through the pairwise spin-spin interactions.  
We carry out computer simulations to examine the efficiency of the model. 
We also show that some chiral representation of the Potts spin enables 
 us to obtain some analytical insights into our labor markets.  
}

\section{Introduction}
\label{sec:Intro}
%
%
Apparently, 
humans (or labors) are the most important resources in our society. 
This is because 
they can produce not only various products and services in the society 
but also they contribute to the society by paying their taxes. 
For this reason, 
in each scale of society, 
say, from nation to companies or much smaller communities such as laboratory (or research group) of university, 
allocation of human resources is one of the essential problems. 
Needless to say, such appropriate allocation of human resource 
is regarded as 
a `matching problem' between 
individuals and some `groups' such as companies, 
and the difference 
among individuals  in their abilities or preference makes the problem 
difficult.

A typical example of the human resource allocation 
is found in simultaneous recruiting of new graduates in Japan. 
Students who are looking for their jobs might research several candidates of companies to enter 
and send the application letter through the web site (what we call `entry sheet'). 
However, the students incline to apply to well-established 
companies, whereas they do not like to get a job in relatively small companies. 
This fact enhances the so-called `mismatch' between labors (students) and companies.  
We can easily see the situation of job-searching process in Japan. 
At the job fair, we find that some booths could collect a lot of students 
(they are all wearing a dark suit even in midsummer!). On the other hand, 
some other booths could not attract the students' attentions. 
Therefore, 
the job-matching itself is apparently governed by some `collective behavior' of 
students. 
Namely, each student seems to behave by looking at their `neighbors' and adapting to 
the `mood' in 
their community, 
or they sometimes can share the useful information 
(of course, such information is sometimes extremely `biased') about the market via Internet or 
social networking service. 

In macroeconomics, there already exist a lot of effective attempts to discuss 
the macroscopic properties \cite{Aoki,Boeri,Roberto,Fagiolo,Neugart} 
including so-called search theory \cite{Lippman,Diamond,Pissarides1985,Pissarides2000}. 
However, apparently, the macroscopic approaches 
lack of their microscopic view points, namely, in their arguments, 
the behavior of microscopic heterogeneous agents 
such as labors or companies are neglected. 

To investigate the collective effects on the job-matching 
process from the microscopic view point, we have proposed several models and 
carried out computer simulations \cite{Chen,Chen2,Chen3} 
by considering some `aggregate data set' for the labor market. 

In our previous successive studies \cite{Chen,Chen2,Chen3}, 
we succeeded in evaluating the macroscopic quantities such as 
unemployment rate $U$ and labor shortage ratio $\Omega$ from the microscopic view point. 
However, these our studies depend on numerical (computer) simulations for relatively small system size to 
calculate these quantities, and 
we definitely need some mathematically rigorous approaches to find  
the universal fact underlying in the job-matching process of labor markets. 
It is also important issue to be considered that 
we should take into account correlation 
between agents (students) 
when we consider the job-matching 
process in realistic labor markets. 
However, in our previous studies \cite{Chen,Chen2,Chen3}, 
we have neglected the correlation in our modeling.

Motivated by the above background and requirement, here we propose 
a mathematical toy model to investigate the job-matching process in 
Japanese labor markets for university graduates and 
investigate the behavior analytically. 
Here we show our preliminary limited results for the typical behavior of the market. 

This paper is organized as follows. 
In section \ref{sec:URN}, we briefly review 
our previous study on the urn model with disorder \cite{Inoue2008} 
and several remarkable properties of the model 
such as Bose-Einstein condensation. 
We also mention that 
the urn model cannot take into account the interactions between 
agents.  
In section \ref{sec:sec1}, we introduce our toy model, the so-called 
Potts model, and 
explain several macroscopic quantities. 
Our preliminary results for 
several job-searching and selection scenarios by students and companies 
are shown in section \ref{sec:sec3}. 
The last section \ref{sec:sec4} is summary and discussion. 
\section{Urn models and Bose condensation: A short review}
\label{sec:URN} 
As a candidate of describing the resource allocation problem, 
we might use the urn models. 
In this model, one can show that a sort of Bose condensation takes place. 
Hence, here we introduce the urn model with a disorder and explain several 
macroscopic properties according to the reference \cite{Inoue2008}. 

We first introduce the Boltzmann weight for the system as 
\begin{equation}
p(\varepsilon_{i}, n_{i})=
\left\{
\begin{array}{cl}
\frac{{\exp}[-\beta E(\varepsilon_{i},n_{i})]}{n_{i}!}
& \mbox{(Each ball is  distinguishable)} \\
{\exp}[-\beta E(\varepsilon_{i},n_{i})] & 
\mbox{(Each ball is  NOT distinguishable)}
\end{array}
\right.
\end{equation}
where $\beta$ stands for the inverse temperature. 
The former is called {\it Ehrenfest class}, 
whereas the latter is referred to as {\it Monkey class}. 

$E(\varepsilon_{i},n_{i})$ denotes 
the energy function for the urn $i$ possessing a disorder $\varepsilon_{i}$ and 
$n_{i}$ balls. 
Obviously, 
in the system with $E(\varepsilon_{i},n_{i}) \propto n_{i} (>0)$, 
each urn (agent) is affected by attractive forces and they 
attempt to gather the balls (resources), 
whereas in the system of $E(\varepsilon_{i},n_{i}) \propto  -n_{i}$, 
each urn is affected by repulsive force and 
they refuse to collect the balls. 
The job-matching process in labor market 
is well-described by the former case. 
On the other hand, 
the problem of spent-nuclear-fuel reprocessing plant in 
Japan is a good example to consider by using 
the latter case, namely, 
balls are `wastes' and urns are `prefectures'.

In the thermodynamic limit: $N,M \to \infty, M/N=\rho=\mathcal{O}(1)$, 
the averaged occupation probability 
$P(k)$, which is a probability that an arbitrary urn possesses $k$ balls is given by 
\[
\rho = 
\left\langle 
\frac{\sum_{n=0}^{\infty}
n \, \phi_{E, \,\mu,\,\beta}(\varepsilon,n)}
{
\sum_{n=0}^{\infty}
\phi_{E, \,\mu,\,\beta}(\varepsilon,n)
}
\right\rangle,\,\,
P(k)=
\left\langle 
\frac{\phi_{E,\,\mu,\,\beta}(\varepsilon,k)}
{\sum_{n=0}^{\infty}
\phi_{E,\,\mu,\,\beta}}
\right\rangle,\,\,\,
z_{s}={\exp}(\beta \mu)
\]
where $z_{s}$ is a solution of the saddle point equation (S.P.E.) and we defined 
\begin{eqnarray}
\phi_{E,\,\mu,\,\beta}(\varepsilon,n)=
\left\{
\begin{array}{ll}
\frac{{\exp}[-\beta (E(\varepsilon,n)-n\mu)]}{n!} & \mbox{(Ehrenfest class)} \\
{\exp}[-\beta (E(\varepsilon,n)-n\mu)] & \mbox{(Monkey class)}
\end{array}
\right.
\end{eqnarray}

In following, we consider the case of Monkey class 
with the cost function: 
\begin{equation}
E(\varepsilon,n)=\varepsilon n, 
\end{equation}
which leads to the 
Boltzmann weight:  
\begin{equation}
\phi_{E,\,\mu,\,\beta}(\varepsilon,n)={\exp}[-\beta n(\varepsilon-\mu)]. 
\end{equation}
We choose the 
distribution of disorder: 
$D(\varepsilon)=\varepsilon_{0} \varepsilon^{\alpha}$. 
Then, the saddle point equation is given by 
\begin{equation}
\rho = 
\int_{0}^{\infty}
\frac{\varepsilon_{0}\varepsilon^{\alpha}d\varepsilon}
{z_{s}^{-1}{\exp}(\beta \varepsilon)-1}
+\rho_{\varepsilon =0} 
\end{equation}
where we should notice that 
$\rho_{\varepsilon =0}$ is negligibly small before condensation. 
We increase the density $\rho$ keeping the temperature $\beta^{-1}$ constant. 
Then, the possible scenario is shown in Table \ref{tab:tb1}. 
\begin{table}
\begin{center}
\begin{tabular}{|c|c|c|}
\hline
density of balls & Solution of S.P.E. &  $\#$ of condensation / $\#$ of non-condensation \\
\hline
$\rho <\rho_{c}$ & $z_{s}<1$ & $0/N\rho$ \\
\hline
$\rho=\rho_{c}$ & $ z_{s}=1$ & $0/N\rho_{c}$ \\
\hline
$\rho >\rho_{c}$ & $z_{s}=1$ & $N(\rho-\rho_{c})/N\rho_{c}$ \\
\hline
\end{tabular}
\end{center}
\caption{\footnotesize The possible scenario of Bose condensation 
controlled by the density $\rho$.}
\label{tab:tb1}
\end{table}
It should be noted that we defined the critical density as 
\begin{equation}
\rho_{c}=\int_{0}^{\infty}
\frac{\varepsilon_{0}\varepsilon^{\alpha}d\varepsilon}{{\exp}(\beta \varepsilon)-1}
\end{equation}

\begin{figure}[ht]
\begin{center}
\includegraphics[width=9cm]{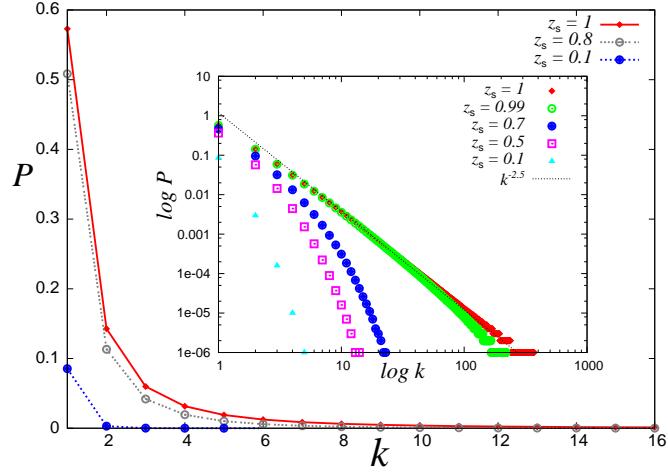}
\end{center}
\caption{\footnotesize 
The occupation probability of the Monkey class urn model 
with disorder.
This figure was taken from 
our previous paper \cite{Inoue2008}.  
}
\label{fig:BE1}
\end{figure}

After simple algebra, we have 
\begin{equation}
P(k) = \frac{z_{s}^{k} \varepsilon_{0}\Gamma(3/2)}{\beta^{3/2}}k^{-3/2}
-
\frac{z_{s}^{k+1}\varepsilon_{0}\Gamma(3/2)}{\beta^{3/2}}
(k+1)^{-3/2}
\end{equation}
for $\alpha=1/2$. 
We show the $P(k)$ for several values of $z_{s}$ in 
Fig.  \ref{fig:BE1}. 
From this figure, we find that 
before condensation, 
namely, 
for $z_{s}<1, \rho <\rho_{c}$, 
the occupation probability is given by 
\begin{equation}
P(k)=\frac{(1-z_{s})\varepsilon_{0}}
{\beta^{3/2}}k^{-3/2}
{\rm e}^{-k\log (1/z_{s})}
\end{equation}
On  the other hand, 
after condensation, 
that is, 
for 
$z_{s}=1,\rho \geq \rho_{c}$, 
we have 
\begin{equation}
P(k)=\frac{3\varepsilon_{0}\Gamma(3/2)}{2\beta^{3/2}}
k^{-5/2}+\frac{1}{N}\delta (k-k_{*})
\end{equation}
The important remarks here are the fact that  
the condensation is specified by the power-law behavior of the occupation probability and 
for the case of without disorder, namely, 
for $D(\varepsilon)=\delta (\varepsilon-\varepsilon_{0})$, 
the power-law behavior disappears. 

As we saw, 
the urn model with disorder exhibits 
a rich physical phenomena such as condensation, however, 
there is no explicit interaction between agents (balls and urns). 
Actually, 
when we consider the job-matching 
process, it is impossible to accept the assumption that 
there is no correlation 
between urns (companies), 
balls (students), or between urns and balls. 
Hence, we should use a different description of the system. 
In the next section, we use the so-called {\it Potts model}  to 
describe the problem of human resource allocation. 
\section{Correlations: The Potts model descriptions}
\label{sec:sec1}
The basic model proposed here for this purpose is
described by the so-called Potts spin glass model which is well-known
in the research field of statistical physics. In the model, each Potts spin
represents the action of each student, and it takes a discrete value (integer) 
corresponding to the company he/she applies for. The pairwise interaction term
in the energy function describes cross-correlations between students, and it makes
our previous model \cite{Chen,Chen2,Chen3} more realistic. 
Obviously, labor science deals with empirical evidence in labor markets and 
it is important for us to look for 
the so-called `stylized facts' which have been discussed mainly in financial markets \cite{Cont,Anirban}. 
We also should reproduce the findings from data-driven models to forecast the market's behavior. 

In following, we show the limited results. Here we consider the system of labor market having $N$ students and $K$ companies. 
To make the problem mathematically tractable, we construct the energy (Hamiltonian) to include
three distinct effects on the students' behavior: 
\begin{equation}
H(\bm{\sigma}_{t}) = - \frac{J}{N}\sum_{ij}c_{ij}\, \delta _{\sigma_{i}^{(t)}, 
\sigma_{j}^{(t)}} -\gamma  \sum_{i = 1}^{N}\sum_{k = 0}^{K -1}\epsilon _{k}\, \delta _{k, \sigma_{i}^{(t)}} 
+ \sum_{i = 1}^{N}\sum_{k = 0}^{K -1}\beta_{k} \left | v_{k}^{*}-v_{k}(t-1)\right | \delta _{k, \sigma_{i}^{(t)}}, 
\label{eq:energy} 
\end{equation}
where $\delta_{a,b}$ denotes a Kronecker's  delta and 
a Potts spin $\sigma_{i}^{(t)}$ stands for the company which student $i$ post his application letter to at stage (or time) $t$, namely, 
\begin{equation}
\sigma_{i}^{(t)} \in \{0,\cdots, K-1\}, \,\,\,i=1,\cdots,N.
\end{equation}
Therefore, the first term in the above equation 
(\ref{eq:energy}) denotes a collective effect, 
the second corresponds to the ranking of companies and 
the third term is a market history. 
In order to include the cross-correlations between students, we describe the system by using the Potts spin glass (see the 
`quenched' random variables $c_{ij}$ in 
(\ref{eq:energy})) as a generalization of the Sherrington-Kirkpatrick model, which is well-known as 
an exactly solvable model for spin glass so far. 
The overall energy function of probabilistic labor market is written explicitly by 
(\ref{eq:energy}).  $c_{ij}$ is an adjacency matrix standing for the `interpersonal relationship' of students, 
and one can choose an arbitrary form, say 
\begin{equation}
c_{ij} =
\left\{
\begin{array}{cl}
c & (\mbox{students $i,j$ are `friendly'}) \\
0 &  (\mbox{students $i,j$ are `independent'}) \\
-c &  (\mbox{students $i,j$ are `anti-friendly'})
\end{array}
\right.
\end{equation}
for $c>0$ 
and the ranking of the company $k$ is defined by $\epsilon _{k}$ (see {\it e.g.} \cite{Chen2} for the detail).  

Before investigating some specific cases below, we shall first provide a general setup. 
Let us introduce a microscopic variable, which represents the 
decision making of companies for a student as 
\begin{equation}
\xi_{i}^{(t)} =
\left\{
\begin{array}{cl}
1 & (\mbox{student $i$ receives an acceptance at stage $t$}) \\
0 & (\mbox{student $i$ is rejected at stage $t$})
\end{array}
\right.
\end{equation}
Then, the conditional probability is given by 
\begin{equation}
P(\xi_{i}^{(t)}|\sigma_{i}^{(t)}) = 
1-A(\sigma_{i}^{(t)})-(1-2A(\sigma_{i}^{(t)}))\xi_{i}^{(t)}
\end{equation}
with the acceptance ratio 
\begin{equation}
A(\sigma_{i}^{(t)})  \equiv  
\sum_{k=0}^{K-1}
\delta_{k,\sigma_{i}^{(t)}}
\Theta (v_{k}^{*}-v_{k}(t))
+
\sum_{k=0}^{K-1}
\delta_{k,\sigma_{i}^{(t)}}\, 
\frac{v_{k}^{*}}{v_{k}(t)}\, 
\Theta (v_{k}(t)-v_{k}^{*}), 
\label{eq:accept0}
\end{equation}
where $v_{k}^{*}(=1/K,\, \mbox{for simplicity in this paper})$ and $v_{k}(t)$ 
denote the quota and actual number of applicants to the company $k$ per student at stage $t$, respectively. $\Theta (\cdots)$ is a conventional step function. 
Hence, when we assume that selecting procedure by companies is 
independent of students, 
we immediately have 
\begin{eqnarray}
P(\bm{\xi}_{t}|\bm{\sigma}_{t}) & = &  
\prod_{i=1}^{N}
P(\xi_{1}^{(t)}|\sigma_{1}^{(t)}) \cdots P(\xi_{N}^{(t)}|\sigma_{N}^{(t)}) \nonumber \\
\mbox{} & = &   
{\exp}
\left[
\sum_{i=1}^{N}
\log \left\{
1-A(\sigma_{i}^{(t)})
-(1-2A(\sigma_{i}^{(t)}))\xi_{i}
\right\}
\right]. 
\end{eqnarray}
Thus, we calculate the joint probability $P(\bm{\xi}_{t},\bm{\sigma}_{t})$ by means of $P(\bm{\xi}_{t}|\bm{\sigma}_{t})P(\bm{\sigma}_{t})$ as 
\begin{eqnarray}
P(\bm{\xi}_{t},\bm{\sigma}_{t}) & = &  
P(\bm{\xi}_{t}|\bm{\sigma}_{t})P(\bm{\sigma}_{t}) \nonumber \\
\mbox{} & = &  
\frac{{\exp}
\left[
\sum_{i=1}^{N}
\log \left\{
1-A(\sigma_{i}^{(t)})
-(1-2A(\sigma_{i}^{(t)}))\xi_{i}^{(t)}
\right\}
-H(\bm{\sigma}_{t})
\right]}
{
\sum_{\bm{\xi}_{t}, \bm{\sigma}_{t}}
{\exp}
\left[
\sum_{i=1}^{N}
\log \left\{
1-A(s_{i}^{(t)})
-(1-2A(s_{i}^{(t)}))\xi_{i}^{(t)}
\right\}
-H(\bm{\sigma}_{t})
\right]
} \nonumber \\
\end{eqnarray}
where 
we assumed that the 
$P(\bm{\sigma}_{t})$ obeys a Gibbs-Boltzmann distribution 
for the energy function (\ref{eq:energy}) as $\sim {\rm e}^{-H(\bm{\sigma}_{t})}$. 

Therefore, the employment rate 
as a macroscopic quantity: 
\begin{equation}
1-U(t) =\frac{1}{N} \sum_{i=1}^{N}\xi_{i}^{(t)}
\end{equation}
is evaluated as an average over the joint probability 
$P(\bm{\xi}_{t},\bm{\sigma}_{t})$, and   
in the thermodynamic limit $N \to \infty$,  it leads to 
\begin{eqnarray}
1-U(t)  & = &   
\frac{
\sum_{\bm{\xi}_{t}, \bm{\sigma}_{t}}
\xi_{i}\,
{\exp}
\left[
\sum_{i=1}^{N}
\log \left\{
1-A(\sigma_{i}^{(t)})
-(1-2A(\sigma_{i}^{(t)}))\xi_{i}^{(t)} 
\right\}
-H(\bm{\sigma}_{t})
\right]}
{
\sum_{\bm{\xi}_{t}, \bm{\sigma}_{t}}
{\exp}
\left[
\sum_{i=1}^{N}
\log \left\{
1-A(\sigma_{i}^{(t)})
-(1-2A(\sigma_{i}^{(t)}))\xi_{i}^{(t)}
\right\}
-H(\bm{\sigma}_{t})
\right]
} \nonumber \\
\mbox{} & = &   
\frac{\sum_{\bm{\sigma}_{t}}
A(\sigma_{i}^{(t)})\, {\exp}[-H(\bm{\sigma}_{t})]}
{\sum_{\bm{\sigma}_{t}}
{\exp}[-H(\bm{\sigma}_{t})]} \equiv 
\langle A(\sigma_{i}^{(t)}) \rangle, 
\label{eq:accept}
\end{eqnarray}
where we defined the bracket: 
\begin{equation}
\langle \cdots \rangle \equiv 
\frac{\sum_{\bm{\sigma}_{t}} 
(\cdots) \, {\exp}[-H(\bm{\sigma}_{t})]}
{\sum_{\bm{\sigma}_{t}}
{\exp}[-H(\bm{\sigma}_{t})]}.
\label{eq:average}
\end{equation}
From the resulting expression (\ref{eq:average}), 
we are confirmed that the employment rate $1-U(t)$ is given by an average of the acceptance ratio (\ref{eq:accept0}) 
over the Gibbs-Boltzmann distribution for the energy function (\ref{eq:energy}). 
Using the above general formula, 
we shall calculate the employment rate exactly for several limited cases. 
\section{The results}
\label{sec:sec3} 
In following, we show our several limited contributions. 
Before we show 
our main result, 
we shall give a relationship between the Potts modeling and 
our previous studies \cite{Chen,Chen2,Chen3} which 
are obtained by simply setting $J=0$ in (\ref{eq:energy}). 
\subsection{For the case of $J=0$}
\begin{figure}[ht]
\begin{center}
\includegraphics[width=5.8cm]{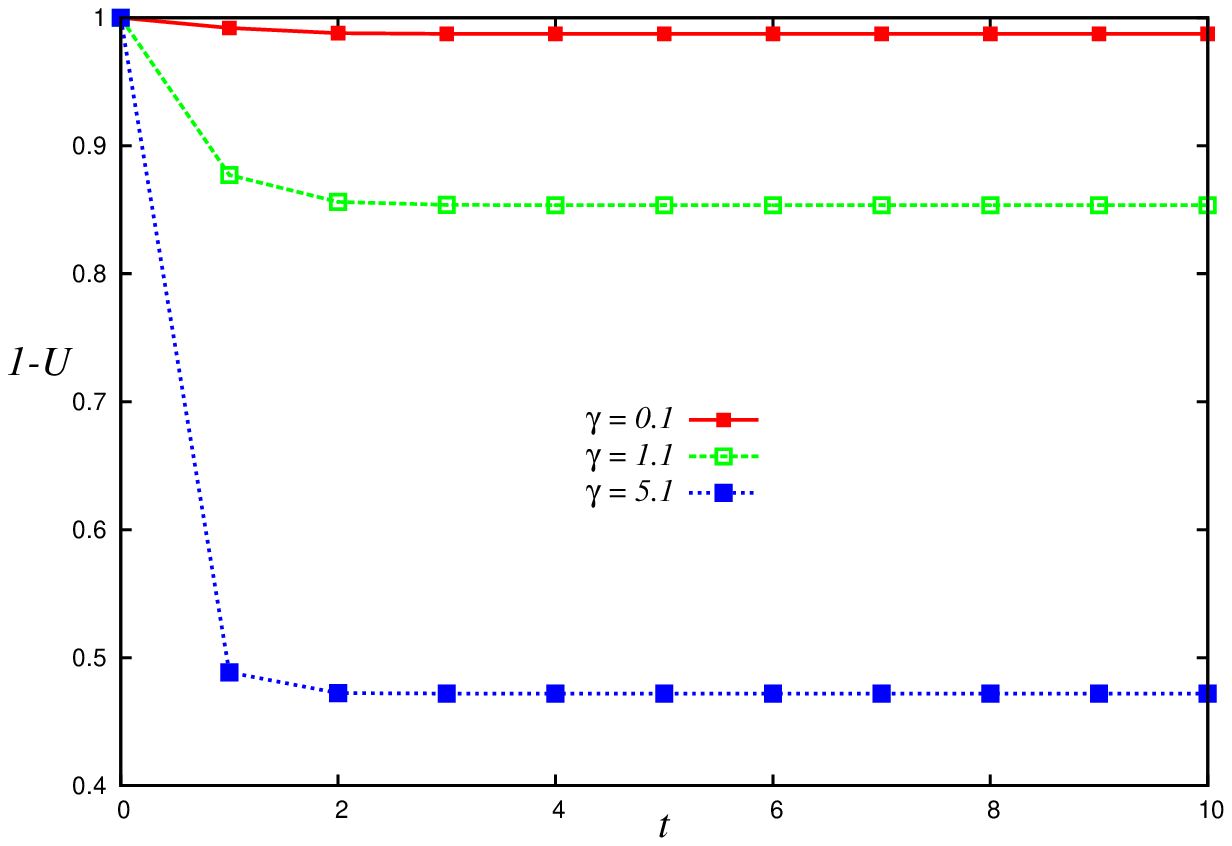} \hspace{-0.2cm}
\includegraphics[width=5.8cm]{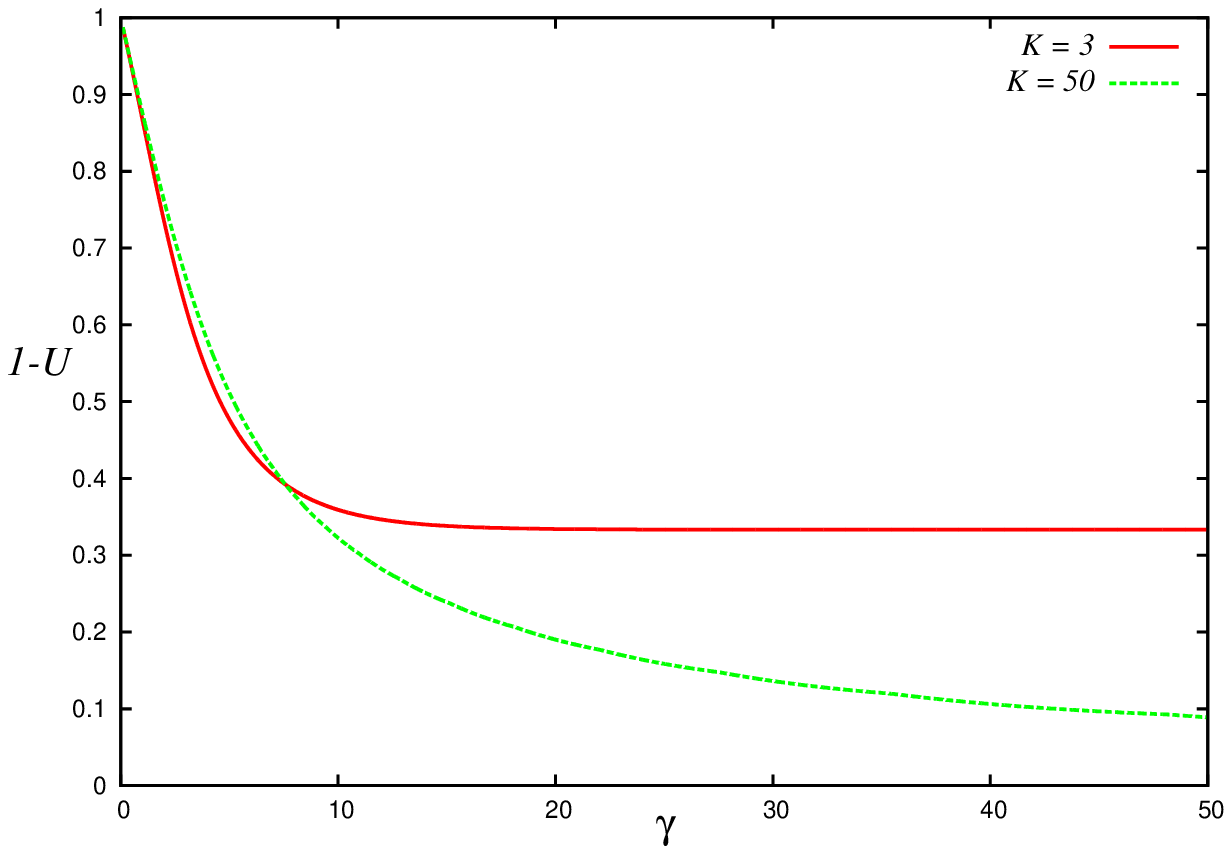} 
\end{center}
\caption{\footnotesize 
The time-dependence of the employment rate for the case of $K=3$ (left) and 
the $\gamma$-dependence of the employment rate 
at the steady state at $t=10$ for $K=3$ and $K=50$ (right). 
We set the job-offer ratio defined in \cite{Chen,Chen2,Chen3} as 
$\alpha=1$ and assume that 
each agent posts only a single application letter to the market, namely, 
$a=1$ in the definition of the previous studies \cite{Chen,Chen2,Chen3}. 
}
\label{fig:fg01}
\end{figure}

We first consider the case of $J=0$. 
For this case, 
the energy function (\ref{eq:energy}) is completely `decoupled'  as follows. 
\begin{eqnarray}
H(\bm{\sigma}_{t})  & = &  \sum_{i}H_{i}, \\
H_{i} & = &   
-\sum_{k=0}^{K-1}
\{
\gamma \epsilon_{k}-
\beta |v_{k}^{*}-v_{k}(t-1)|
\}\delta_{\sigma_{i}^{(t)},k}
\end{eqnarray}
where we set $\beta_{k}=\beta\,(\forall_{k})$ for simplicity. 
Hence, the 
$v_{k}(t)$ is evaluated in terms of the definition (\ref{eq:average}) as
\begin{equation}
v_{k}(t)  \equiv 
\lim_{N \to \infty}   
\frac{1}{N} 
\sum_{i=1}^{N}\delta_{\sigma_{i}^{(t)},k} =  
\left\langle \delta_{\sigma_{i}^{(t)},k} \right\rangle = 
\frac{{\exp} [-\gamma\epsilon_{k}+\beta |v_{k}^{*}-v_{k}(t-1)|]}
{\sum_{k=0}^{K-1}
{\exp} [-\gamma\epsilon_{k}+\beta |v_{k}^{*}-v_{k}(t-1)|]}
\label{eq:evol_v}
\end{equation}
and from the expression of employment rate (\ref{eq:accept}), we have 
\begin{equation}
1-U(t) =
\frac{
\sum_{k=0}^{K-1}
\left\{
\frac{v_{k}^{*}}
{v_{k}(t)}
+
\left(
1-
\frac{v_{k}^{*}}{v_{k}(t)}
\right)
\Theta (v_{k}^{*}-v_{k}(t))
\right\}{\exp} [-\gamma\epsilon_{k}+\beta |v_{k}^{*}-v_{k}(t-1)|]
}
{\sum_{k=0}^{K-1}
{\exp} [-\gamma\epsilon_{k}+\beta |v_{k}^{*}-v_{k}(t-1)|]}. 
\label{eq:1-U}
\end{equation}
By solving the non-linear equation 
(\ref{eq:evol_v}) recursively and substituting the solution 
$v_{k}(t)$ into (\ref{eq:1-U}), we obtain the time-dependence of 
the employment rate $1-U(t)$. 
In Fig. \ref{fig:fg01}, 
we plot the time-dependence of the employment rate for the case of $K=3$ (left) and 
the $\gamma$-dependence of the employment rate 
at the steady state at $t=10$ for $K=3$ and $K=50$ (right). 
We set the job-offer ratio defined in \cite{Chen,Chen2,Chen3} as 
$\alpha=1$. 
The ranking factor is also selected by 
\begin{equation}
\epsilon_{k}=1+ \frac{k}{K}. 
\end{equation}
We here assumed that 
each agent posts only a single application letter to the market, namely, 
$a=1$ in the definition of the previous studies \cite{Chen,Chen2,Chen3}. 
It should be important for us to remind that 
the above equation 
(\ref{eq:evol_v}) is exactly the same as the update rule for the aggregation probability 
$P_{k}(t)$ in the reference \cite{Chen}. 
However, when we restrict ourselves to the case of 
$\alpha=a=1$, one can obtain 
the time-dependence of the employment rate 
exactly by (\ref{eq:1-U}). 
This is an advantage of this approach. 
It also should be noted that from the relationship: 
\begin{equation}
U=\alpha \Omega + 1-\alpha
\label{eq:UOmega}
\end{equation}
(see \cite{Chen} for the derivation), 
we have $U=\Omega$, namely, the unemployment rate is exactly the same as 
the labor shortage ratio for $\alpha=1$. 

It is important for us to notice that 
the aggregation probability of 
the system $P(\bm{\sigma}_{t})$ is 
rewritten in terms of $P_{k}(t)$ in the references \cite{Chen,Chen2,Chen3} as 
\begin{equation}
P(\bm{\sigma}_{t}) = 
\left\{
\prod_{k=1}^{K}P_{k}(t)
\right\}^{N}
\end{equation}
with $P_{k}(t)=v_{k}(t)$ (see (\ref{eq:evol_v})) even for $\alpha \neq 1$. 
For this case, 
the system parameters are only $\gamma$ and $\beta$, 
and these unknown parameters are easily calibrated from the empirical data \cite{Chen3}. 
As the result, we obtained $U$-$\Omega$ curve using (\ref{eq:UOmega})  
for the past 17 years in Japanese labor market for university graduates. 
\begin{figure}[ht]
\begin{center}
\includegraphics[width=9cm]{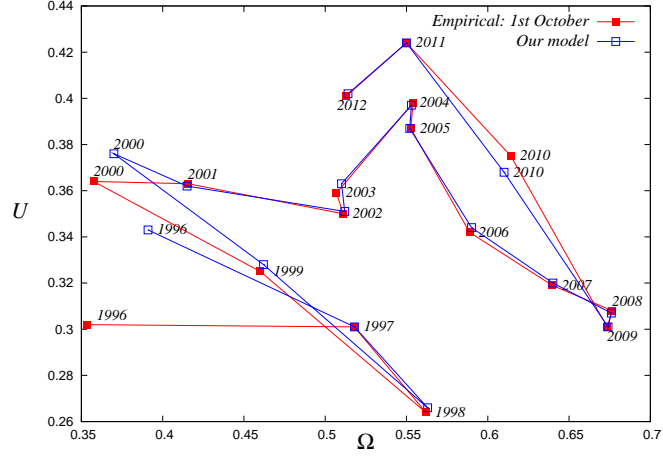}
\end{center}
\caption{\footnotesize 
The empirical and theoretical $U$-$\Omega$ curves. 
We clearly find that the large $\gamma$ apparently pushes 
the $U$-$\Omega$ curve toward the upper right direction where 
the global mismatch between the students and companies is large. 
The picture was taken from our previous study \cite{Chen3}. 
}
\label{fig:fg_additional}
\end{figure}
We plot the resulting and $U$-$\Omega$ curve in Fig. \ref{fig:fg_additional}. 
The gap between the theoretical and empirical curves 
comes from the uncertainties in the calibration of average number of application letters $a$. 
In this figure, we simply
chose the value as $a=10$ in our calculations. 
\subsection{The case of $J \neq 0$}
We next  consider the case of 
$J \neq 0$. 
Then, we should note that some `chiral representation' of 
the energy function (\ref{eq:energy}) by means of 
the chiral Potts spin \cite{Nishimori,Carlucci} (Note: `$i$' appearing in `$2\pi i$' below is an imaginary unit): 
\begin{equation}
\lambda_{i} = {\exp}
\left(
\frac{2\pi i}{K} \sigma_{i}^{(t)}
\right), \, \sigma_{i}^{(t)}=0,\cdots,K-1
\label{eq:chiral}
\end{equation}
enables us to obtain some analytical insights into our labor markets. 
\subsubsection{The case of $\gamma=\beta=0$: Without ranking and market history}
\begin{figure}[ht]
\begin{center}
\includegraphics[width=5.8cm]{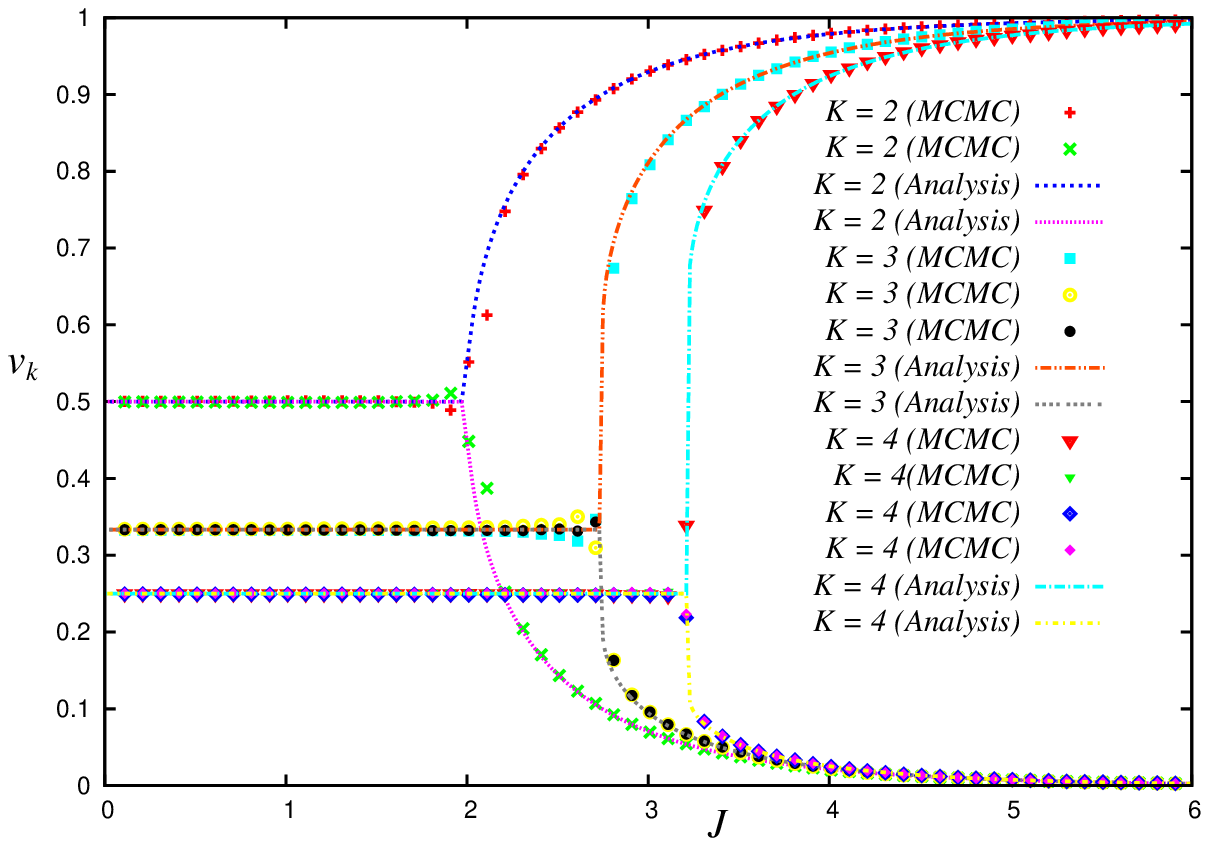} \hspace{-0.2cm}
\includegraphics[width=5.8cm]{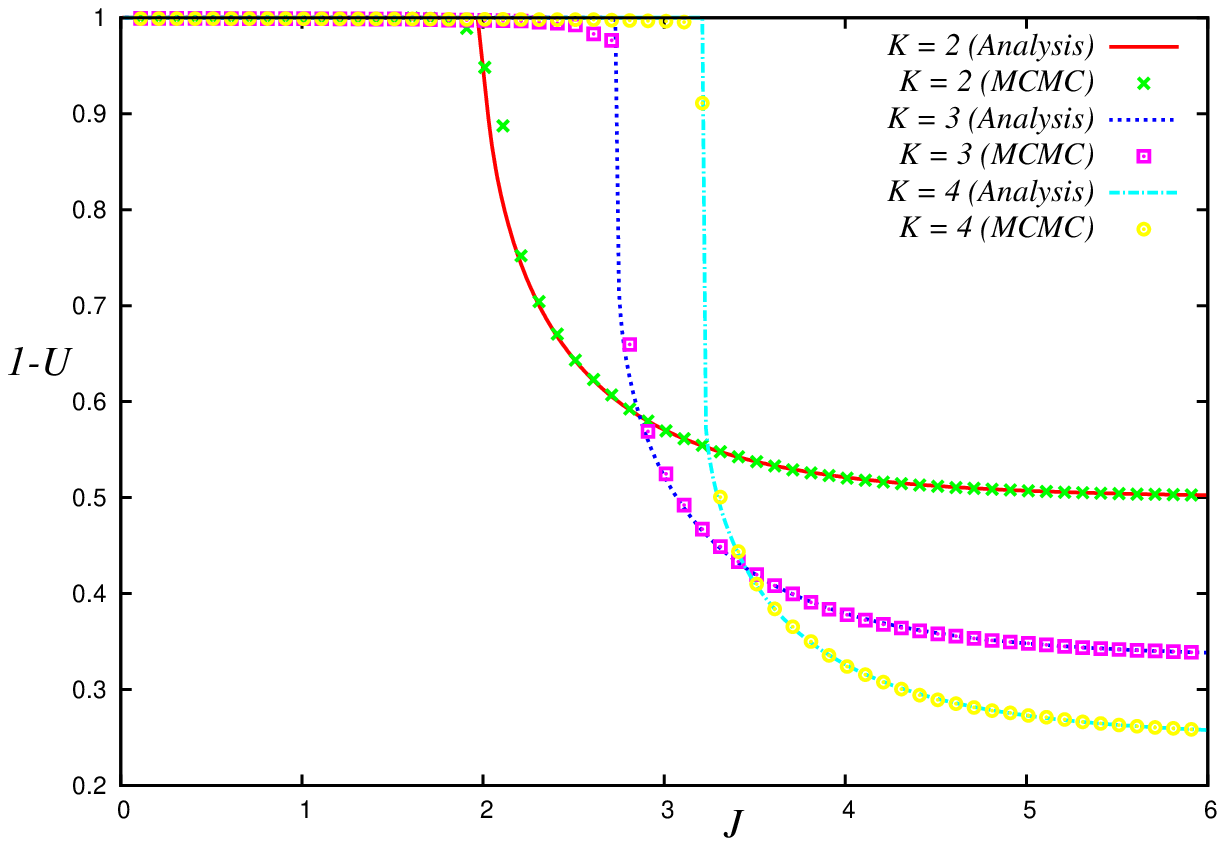}
\end{center}
\caption{\footnotesize  
The actual number of applicants $v_{k}$ (left) and employment rate $1-U$ (right) as a function of the strength of cooperation $J$. 
We find that the system undergoes a phase transition at the critical point. 
The transition is the second order for $K=2$, whereas it is the first order for $K \geq 3$. 
These critical points are given by 
$J_{\rm c}=2$ for $K=2$, 
$J_{\rm c}=2.73$ for $K=3$ and $J_{\rm c}=3.21$ for $K=4$. 
We should mention that 
the analytic results (lines) and the corresponding Monte Carlo simulations (MCMC) 
with the finite number of  students $N=1000$  (dots) 
are in an excellent agreement.  
We should notice that perfect employment phase is a `disordered phase', 
whereas the poor employment phase corresponds to an `ordered phase' in the literature of 
order-disorder phase transition. 
For large strength of cooperation $J$, as a company occupies all applications up to the quota, 
$\lim_{J \to \infty}(1-U)=v_{k}^{*}=1/K$ (the quota per student) is satisfied. 
}
\label{fig:fg1}
\end{figure}

As a preliminary, we show the employment rate $1-U$ as a function of $J(>0)$ for the simplest case 
$\gamma=\beta=0$ and $c_{ij}=1\, (\forall_{ij})$ in Fig.  \ref{fig:fg1} (right),  and 
the actual number of applicants the company $k$ obtains in Fig. \ref{fig:fg1} (left).  
We should keep in mind that for this simplest case with local energy  
\begin{equation}
H_{ij}  \equiv  -J \delta_{\sigma_{i},\sigma_{j}} = 
-\frac{J}{K}\sum_{r=0}^{K-1}
\lambda_{i}^{r} \lambda_{j}^{K-r} = -\frac{J}{K} \left\{
1+ 
\sum_{r=1}^{K-1} 
\lambda_{i}^{r} \lambda_{j}^{K-r}
\right\}
\end{equation}
under the transformation (\ref{eq:chiral}) leading to the total energy $H(\bm{\sigma}) \equiv \sum_{ij}H_{ij}$, by evaluating the partition function: 
\begin{equation}
Z  =   
\sum_{\bm{\sigma}}
{\exp}
\left[
\frac{J}{NK}
\sum_{r=1}^{K-1}
\sum_{ij}
\cos 
\frac{2\pi r (\sigma_{i}-\sigma_{j})}{K}
\right] 
\end{equation}
in the limit of $N \to \infty$, 
one can obtain the employment rate 
$1-U=\langle A(\bm{\sigma}) \rangle$ (see also equation (\ref{eq:accept})) exactly as 
\begin{eqnarray}
1- U & = &    \frac{\sum_{\bm{\sigma}}
A(\bm{\sigma})\, {\exp}[-H(\bm{\sigma)}]}
{\sum_{\bm{\sigma}}
{\exp}[-H(\bm{\sigma})]} \nonumber \\
\mbox{} & = & 
\frac{
\left\{
\frac{v_{0}^{*}}{v_{0}}+ 
\left(
1-\frac{v_{0}^{*}}{v_{0}}
\right) \Theta (v_{0}^{*}-v_{0})
\right\}
}
{1 + 
(K-1){\rm e}^{-\frac{Jx}{K-1}}
}  + 
\frac{
(K-1)
\left\{
\frac{v_{k}^{*}}{v_{k}}+ 
\left(
1-\frac{v_{k}^{*}}{v_{k}}
\right) \Theta (v_{k}^{*}-v_{k})
\right\}
{\rm e}^{-\frac{Jx}{K-1}}}
{1+ 
(K-1){\rm e}^{-\frac{Jx}{K-1}}
} \nonumber \\
\end{eqnarray}
with 
\begin{equation}
v_{k} \equiv   \lim_{N \to \infty} \frac{1}{N}\sum_{i=1}^{N}\delta_{\sigma_{i},k} = 
\left\langle \delta_{\sigma,k} \right\rangle = 
\frac{\delta_{0,k} + 
\sum_{\sigma=1}^{K-1} \delta_{\sigma,k}\,{\rm e}^{-\frac{Jx}{K-1}}}
{1+(K-1)\, {\rm e}^{-\frac{Jx}{K-1}}},\,\,\,k=0,\cdots, K-1, 
\end{equation}
where an order parameter $x$ is determined as a solution of the following non-linear equation: 
\begin{equation}
x = (K-1)
\left(
\frac{1-{\rm e}^{-\frac{J}{K-1}x}}
{1+(K-1){\rm e}^{-\frac{J}{K-1}x}}
\right). 
\label{eq:sol_v}
\end{equation}
It should be noted that the above $x$ is given by the extremum of the free energy density: 
\begin{equation}
f = 
-\frac{Jx^{2}}{K(K-1)}  + \log 
\sum_{\sigma=0}^{K-1}
{\exp}
\left[
\frac{Jx}{K(K-1)}
\sum_{r=1}^{K-1}
\cos 
\left(\frac{2\pi r}{K} \sigma
\right)
\right]. 
\end{equation}
The acceptance ratio $A(\bm{\sigma})$ is now given by 
\begin{equation}
A(\bm{\sigma}) \equiv \sum_{i=1}^{N} A(\sigma_{i}) = 
\sum_{i=1}^{N} \sum_{k=0}^{K-1} \delta_{\sigma_{i},k}
\left\{
\frac{v_{k}^{*}}{v_{k}}+ 
\left(
1-\frac{v_{k}^{*}}{v_{k}}
\right) \Theta (v_{k}^{*}-v_{k})
\right\},
\end{equation} 
and we omitted the time $t$-dependence in the above expressions 
because the system is no longer dependent on the market history, 
namely $v_{k}(t-1)$,  for the choice of $\beta=\gamma=0$ 
in the energy function (\ref{eq:energy}). 

In Fig.  \ref{fig:fg1}, we easily find that phase transitions take place when the strength of `cooperation' $J$ increases beyond the critical point $J_{\rm c}$. 
Namely, for weak $J$ regime, 
`random search' by students is a good strategy to realize 
the perfect employment state ($1-U=1$), however, once $J$ increases beyond the critical point, 
the perfect state is no longer stable and 
system suddenly goes into the extremely worse employment phase for $K \geq 3$ (first order phase transition). 
The critical point of 
the second order phase transition for $K=2$ is 
easily obtained by expanding (\ref{eq:sol_v}) around $x=0$ as 
\begin{equation}
x = \frac{1-{\rm e}^{-Jx}}{1+{\rm e}^{-Jx}} \simeq Jx/2
\end{equation}
and this reads $J_{\rm c}=2$. 
For the first order phase transition, we numerically 
obtain the critical values, 
for instance, we have 
$J_{\rm c}=2.73$ for $K=3$ and $J_{\rm c}=3.21$ for $K=4$. 
As the number $K$ is quite large far beyond $K=3$ in real labor markets, hence the above finding for the discontinuous 
transition might be useful for discussing 
a mismatch between students and companies, which is a serious issue in 
recent Japanese labor markets (see the reference \cite{Chen3}). 
   
We also carried out computer simulations to examine
the efficiency of the model.  We should mention that 
the analytic results (lines) and the corresponding Monte Carlo simulations (dots) 
with finite system size $N=1000$ are in an excellent agreement in the figures. 
This preliminary result is a justification for us to conform that 
one can make a mathematically rigorous platform to investigate 
the labor market along this direction. 
\mbox{}

We next consider the case of 
$\beta,\gamma \neq 0$. 
\subsubsection{Ranking effects}
For the case of $\gamma \neq 0, \beta_{k}=0\,(\forall_{k})$, 
the saddle point equation is given by the following two-dimensional vector form: 
\begin{equation}
(x_{r},y_{r})=\langle \bm{u}_{r}(s) \rangle_{*}=\left(
\left\langle \cos \frac{2\pi r}{K}s \right\rangle_{*}, 
\left\langle \sin \frac{2\pi r}{K}s \right\rangle_{*}
\right),\,\,\,r=0,\cdots,K-1
\label{eq:vector}
\end{equation}
where we defined the bracket $\langle \cdots \rangle_{*}$ as 
\begin{eqnarray}
&& \langle \cdots \rangle_{*}  \equiv 
\frac{\sum_{s=0}^{K-1}(\cdots) \,{\exp}[\psi_{r}(s: \{x_{r}\},\{y_{r}\})]}
{\sum_{s=0}^{K-1} {\exp}[\psi_{r}(s: \{x_{r}\}, \{y_{r}\})]}, \\
&& \psi_{r}(s: \{x_{r}\},\{y_{r}\}) \equiv  \sum_{r=0}^{K-1}\bm{X}_{r} \cdot \bm{u}_{r}(s)
\end{eqnarray}
with the following two vectors: 
\begin{eqnarray}
\bm{X}_{r} & = & \left(
\frac{J}{K}x_{r}+ \frac{\gamma}{K} \sum_{k=0}^{K-1}
\epsilon_{k}\cos \frac{2\pi r}{K}k,
\frac{J}{K}y_{r}+ \frac{\gamma}{K} \sum_{k=0}^{K-1}
\epsilon_{k}\sin \frac{2\pi r}{K}k
\right) \\
\bm{u}_{r}(s) & = &  
\left(
\cos \frac{2\pi r}{K}s,\sin\frac{2\pi r}{K}s
\right). 
\end{eqnarray}
From the energy function (\ref{eq:energy}) and the above formula, 
we  should notice that the ranking factor 
$\epsilon_{k}$ is regarded as a `state-dependent field' affecting 
each spin and 
the symmetry in the `perfect employment phase' for small $J$ (see Fig. \ref{fig:fg1} (right)) might be broken by these unbiased effects.  
We also should keep in mind that 
for the case of $\gamma=0$ or 
$\epsilon_{k}=\epsilon\,\,(\forall_{k})$, 
we find that the equation (\ref{eq:vector}) possesses  
the solution of the type: $x_{0},\cdots,x_{K-1}\neq 0,\, y_{0}=\cdots=y_{K-1}=0$. 
It should be also bear in mind that 
$K=2$ is rather a special case and the solution of the above type is obtained simply as 
\begin{equation}
x_{0}=1,\, x \equiv x_{1}=
\frac{1-{\rm e}^{-Jx+\gamma (\epsilon_{1}-\epsilon_{0})}}
{1+{\rm e}^{-Jx+\gamma (\epsilon_{1}-\epsilon_{0})}},\,\,
y_{0}=y_{1}=0. 
\end{equation}
However, for general case, we must deal with two-dimensional vectors $(x_{r},y_{y}),\,r=0,\cdots,K-1$ 
with each non-zero component $x_{r},y_{r} \neq 0$ 
to specify the equilibrium properties of the system. 

\begin{figure}[ht]
\begin{center}
\includegraphics[width=5.8cm]{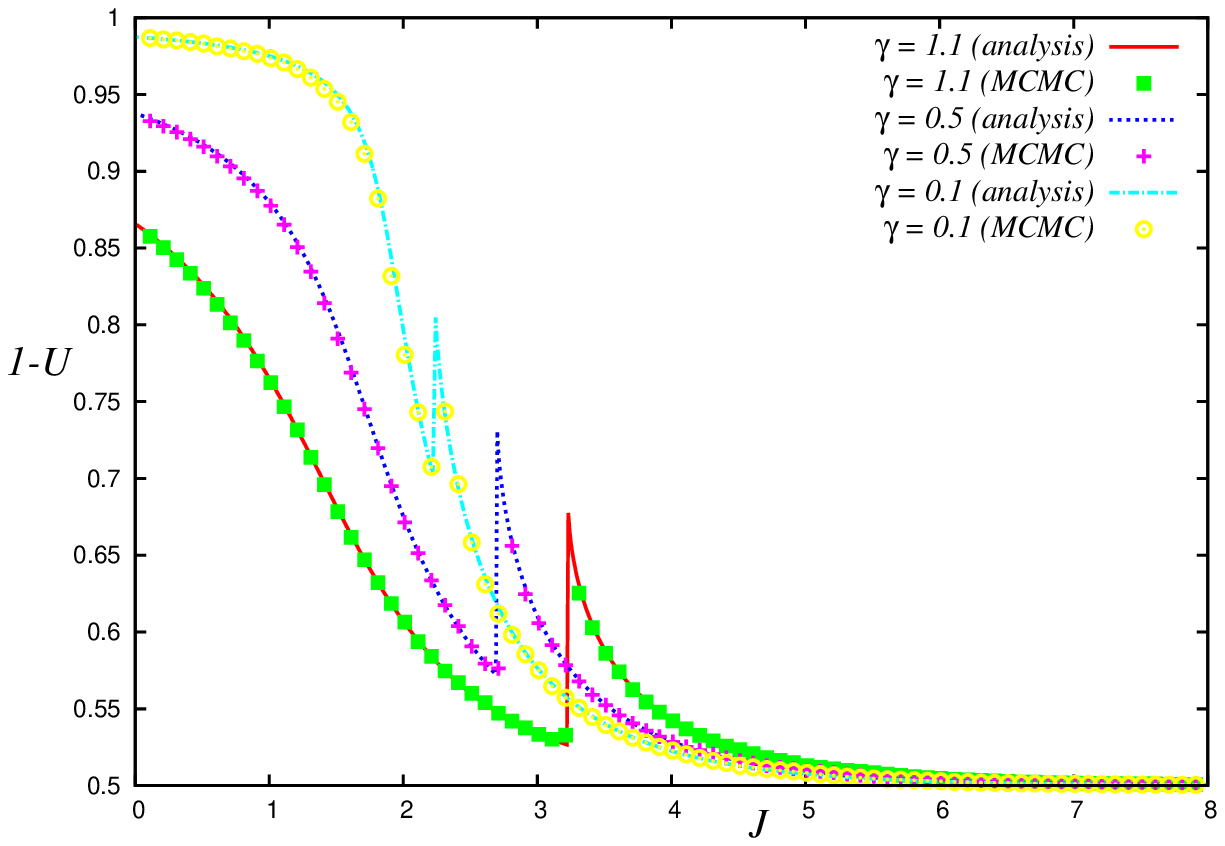} \hspace{-0.2cm}
\includegraphics[width=5.8cm]{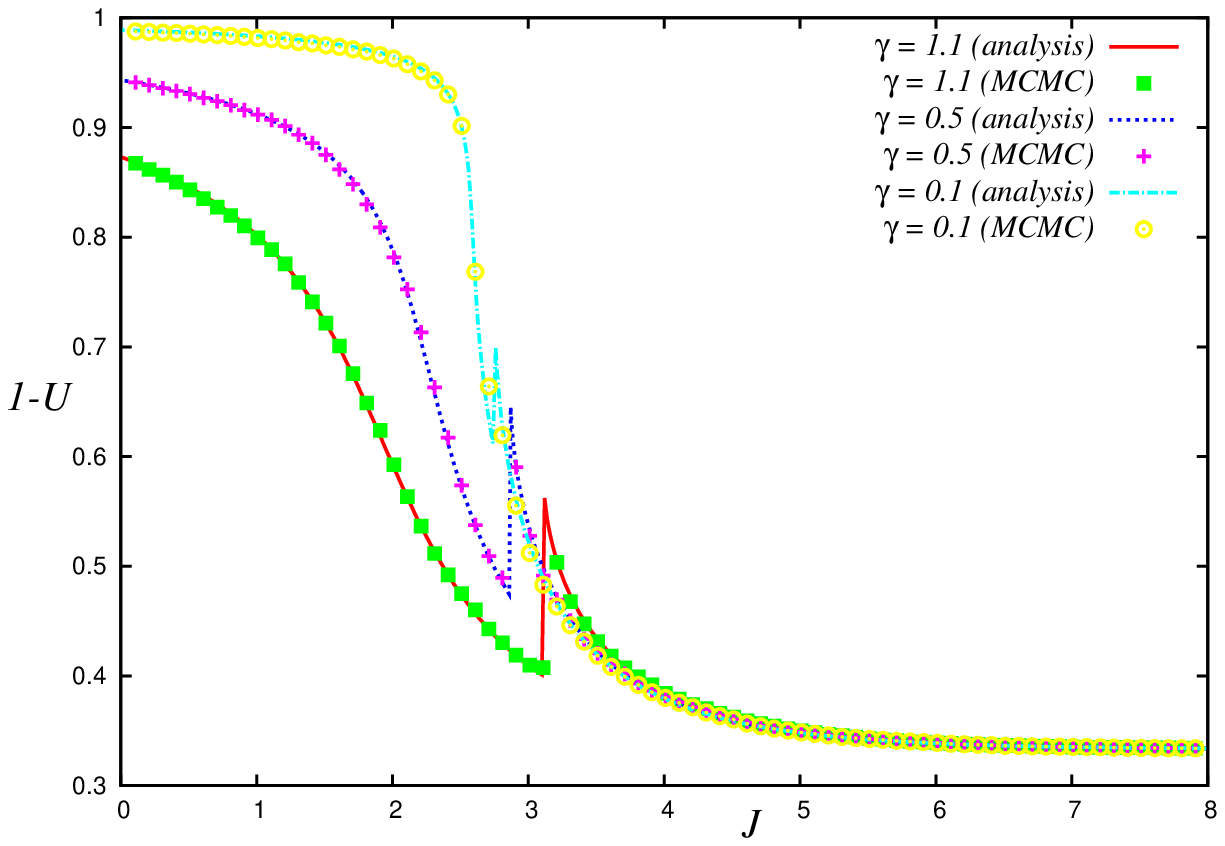}
\end{center}
\caption{\footnotesize  
The strength of cooperation $J$-dependence of 
the employment rate for the case of 
$\gamma \neq 0, \beta_{k}=0\,(\forall_{k})$. 
We plot the case of $K=2$ (left) and $K=3$ (right). 
We find that the phase transition as shown in 
Fig.  \ref{fig:fg1} 
disappears, however, 
the ergodicity breaking phase 
appears within intermediate range of $J$. 
We are conformed that 
$\lim_{J \to \infty}(1-U)=1/K$ is satisfied even for this case. 
The simulations (MCMC) are carried out for the system of size $N=1000$. 
}
\label{fig:fg2}
\end{figure}

For the solution $(x_{r},y_{r}),\,r=0,\cdots,K-1$, 
we obtain the order parameters and employment rate as 
\begin{eqnarray}
v_{r} & = & \langle \delta_{r,s} \rangle_{*} \\
1-U & = &  \langle A(s) \rangle_{*},\,\,\, r=0,\cdots,K-1. 
\end{eqnarray}
In Fig.  \ref{fig:fg2}, we plot 
the $J$-dependence of 
the employment rate for $K=2$ (left) and $K=3$ (right). 
From this figure, we find that 
the employment rate decreases monotonically, however, 
within intermediate range of 
$J$, the $1-U$ behaves discontinuously. 
We should notice that 
in this regime, 
the `ergodicity' of the system might be broken 
because 
the realized value of $1-U$ by Monte Carlo simulation 
is strongly dependent on the choice of initial configuration (pattern) of 
Potts spins. 

\begin{figure}[ht]
\begin{center}
\includegraphics[width=9cm]{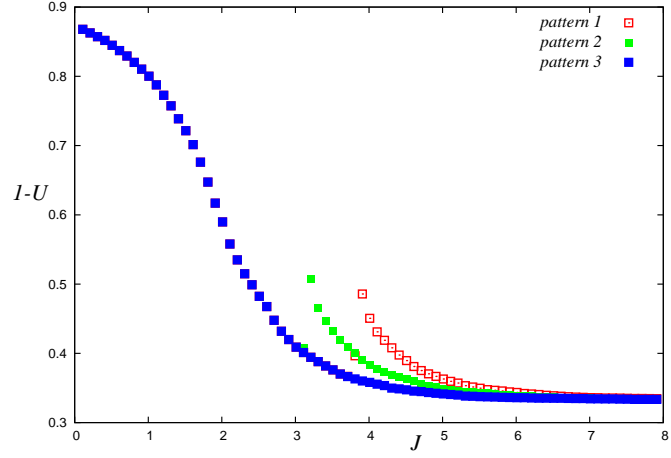} 
\end{center}
\caption{\footnotesize  
The initial configuration dependence of 
the $1-U$. We set $K=3, \gamma=1.1$ and 
choose three distinct initial configurations for 
Monte Carlo simulations. We find that 
$1-U$ is strongly dependent on 
the initial condition (`{\tt pattern $1\sim 3$}') within intermediate range of $J$. 
In this plot, we chose 
the two distinct initial conditions so as to make the 
gap of order parameters $\mathcal{O}(1)$ object, 
that is, 
$\Delta x_{r} (\equiv x_{r}^{(\rm a)}-x_{r}^{({\rm b})}), 
\Delta y_{r} (\equiv y_{r}^{(\rm a)}-y_{r}^{({\rm b})}) \sim \mathcal{O}(1)$ 
for $r=0,\cdots, K-1$ $({\rm a,b} =\{\mbox{\tt pattern 1,pattern 2, pattern 3}\})$. 
}
\label{fig:fg22}
\end{figure}

To see the result more explicitly, 
we should draw our attention to the initial condition dependence 
of the $J$-$(1-U)$ curve.  
Actually, here we carry out Monte Carlo simulation 
to examine the initial configuration dependence of 
the $1-U$ numerically and show the results in Fig. \ref{fig:fg22}. 
From this figure, we confirm that 
the value of 
the $1-U$ depends on the initial configuration of 
the Potts spins although 
the $1-U$ is independent of the initial condition 
for $J<3$ and $J \gg1$. 
In this plot, we chose 
the two distinct initial conditions so as to make the 
gap of order parameters $\mathcal{O}(1)$ object, 
that is, 
\begin{equation}
\Delta x_{r} (\equiv x_{r}^{(\rm a)}-x_{r}^{({\rm b})}), 
\Delta y_{r} (\equiv y_{r}^{(\rm a)}-y_{r}^{({\rm b})}) \sim \mathcal{O}(1)
\end{equation}
for $r=0,\cdots, K-1$. 
\mbox{}

It might be important for us to 
investigate the basin of attraction for the matching dynamics 
analytically as in the reference \cite{Inoue}, however, 
it is far beyond the scope of the current paper and it should be addressed our future study. 
\mbox{}
\subsubsection{Market history effects}
We next consider the case of 
$\beta_{k} \neq 0 \,(\forall_{k})$.  
For this case, we should replace 
the $\bm{X}_{r}$ in the saddle point equation (\ref{eq:vector}) by 
\begin{eqnarray}
\bm{X}_{r} & = & {\Biggr (}
\frac{J}{K}x_{r}+ \frac{1}{K} \sum_{k=0}^{K-1}
(\gamma \epsilon_{k}- 
\beta_{k}|v_{k}^{*}-v_{k}(t-1)|)\cos \frac{2\pi r}{K}k, \nonumber \\
\mbox{} 
\mbox{} && \frac{J}{K}y_{r}+ \frac{1}{K} \sum_{k=0}^{K-1}
(\gamma \epsilon_{k}- 
\beta_{k} |v_{k}^{*}-v_{k}(t-1)|)\sin \frac{2\pi r}{K}k
{\Biggr )}.
\label{eq:Xr}
\end{eqnarray}
It should be noticed 
that 
the $v_{k}$ at the previous stage $t-1$ 
is regarded as an `external field' which affects 
the spin system 
at the current stage $t$. 
Hence, 
by substituting $v_{k}(0)$ as an initial state into 
the equation (\ref{eq:vector}) with (\ref{eq:Xr}), 
we can solve the equation with respect to $v_{k}(1)$.  
By repeating the procedures, we obtain the `time series' as 
$v_{k}(0) \to v_{k}(1) \to \cdots v_{k}(t) \to $ for all $k$ and 
$1-U(t)$ as a function of $t$. 
\begin{figure}[ht]
\begin{center}
\includegraphics[width=5.8cm]{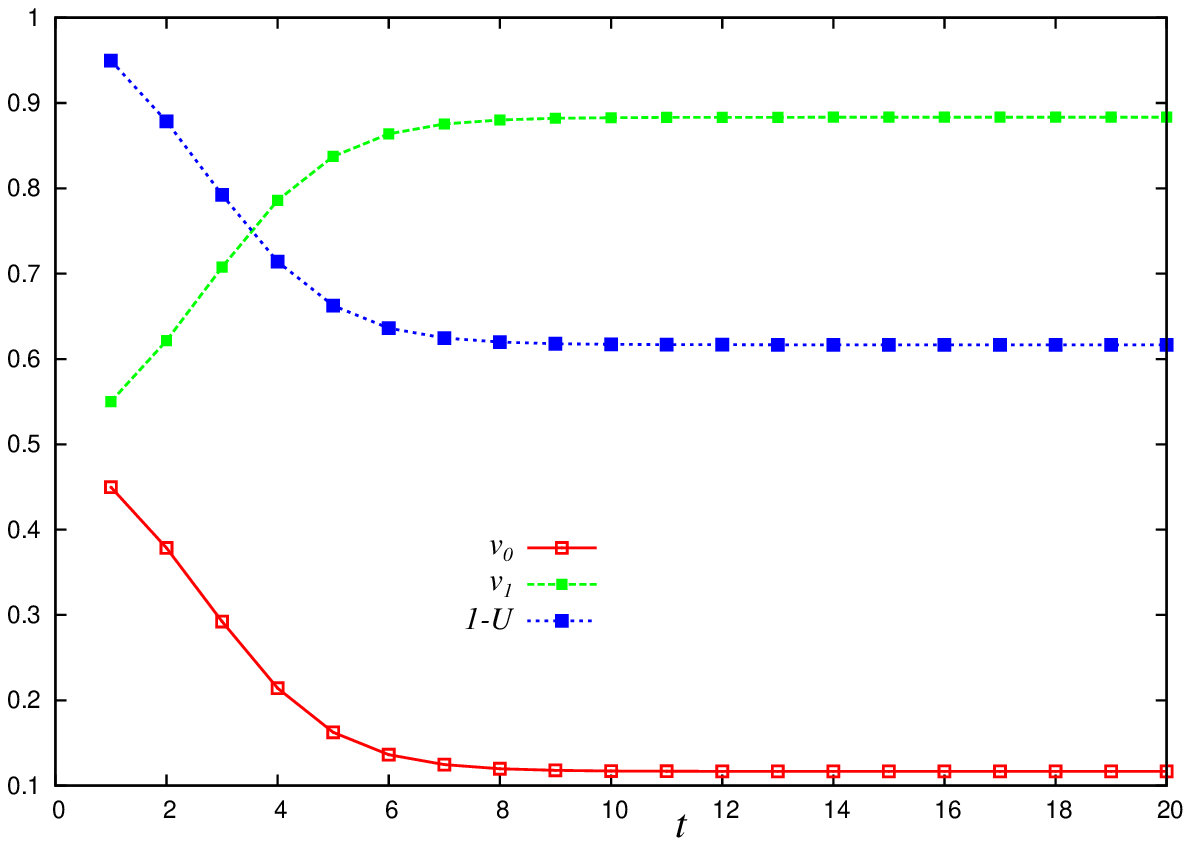} \hspace{-0.2cm}
\includegraphics[width=5.8cm]{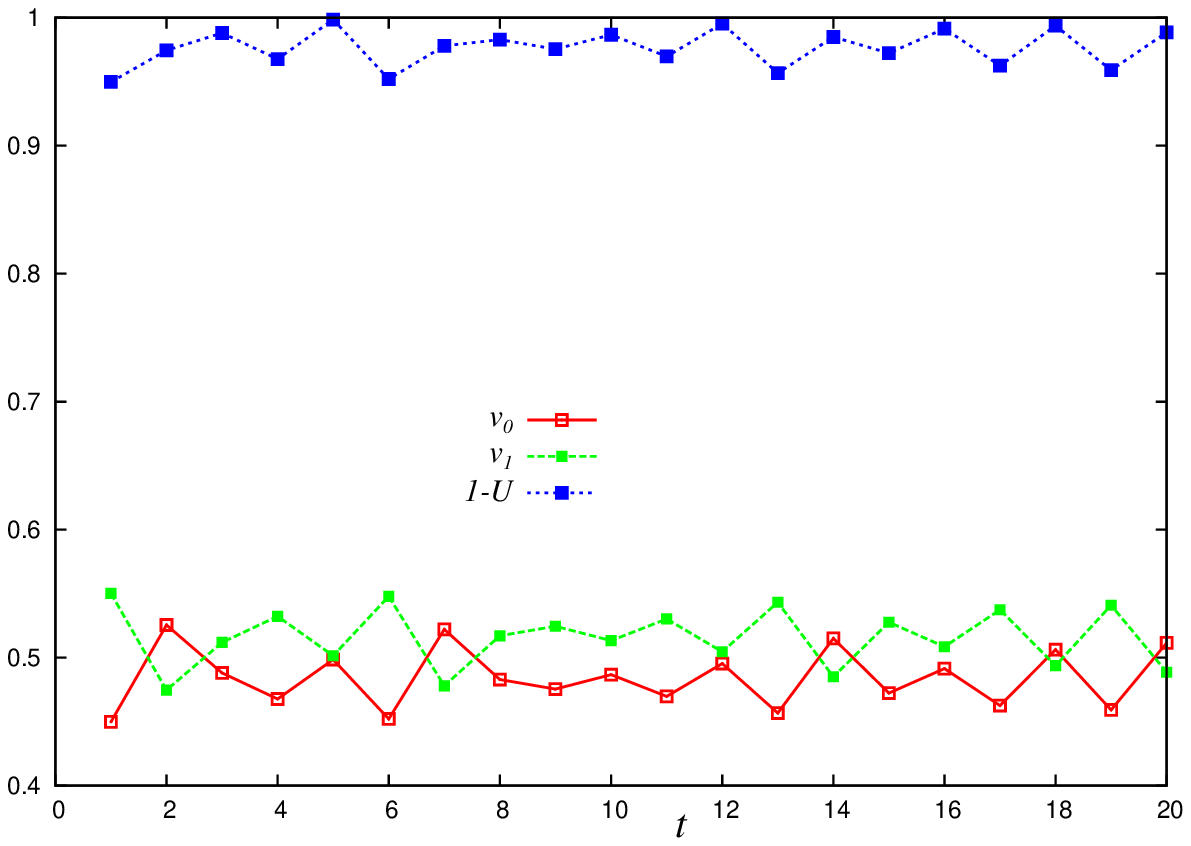}
\end{center}
\caption{\footnotesize  
The time (stage) dependence of 
the employment rate $1-U$ for the case of 
$K=2,J=1,\gamma=0.1$ and $(\beta_{1},\beta_{0})=(1,4)$ (left) and 
$(\beta_{1},\beta_{0})=(4,1)$ (right). 
The `zigzag behavior' 
in $v_{k}(t)$ is observed for $(\beta_{1},\beta_{0})=(4,1)$. 
}
\label{fig:fg3}
\end{figure}
In Fig.  \ref{fig:fg3}, we 
plot the time (stage) dependence of 
the employment rate $1-U$ for the case of 
$K=2,J=1,\gamma=0.1$ and $(\beta_{1},\beta_{0})=(1,4)$ (left) and 
$(\beta_{1},\beta_{0})=(4,1)$ (right). 
From this figure, 
we find that 
the larger weight of 
the market history effect for the highest ranking company $\beta_{1}$ in comparison with $\beta_{0}$ 
induces the periodical change of the order for $v_{1},v_{0}$ due to 
the negative feedback (a sort of `minority game' \cite{Minority} for the students). 
Namely, 
from the ranking gap $\epsilon_{1}-\epsilon_{0}=1/2$ for $K=2$, 
the company `1' attracts a lot of 
applications at time $t$ even for a relatively small strength of the preference $\gamma=0.1$. 
However, 
at the next stage, 
the ability of the aggregation for 
the company `1' remarkably decreases due to the 
large $\beta_{1}$. As the result, 
the inequality $v_{1}>v_{0}$ is reversed as $v_{0}>v_{1}$, and 
the company `0' obtains much more applications than the company `1'  at this stage. 
After several time steps, 
the amount of $\beta_{1}|v_{1}^{*}-v_{1}(t-1)|$ 
becomes small enough to turn on the switch of the preference for 
the high ranking company `1', and eventually the inequality $v_{1}>v_{0}$ should be recovered again.  
The `zigzag behavior' due to the above feedback mechanism 
in $v_{k}(t)$ is actually observed in Fig. \ref{fig:fg3} (right). 
On the other hand, 
when 
the strength of the history effect $\beta_{0}$ for the lower ranking company is larger than that of 
the higher ranking company $\beta_{1}$, 
the zigzag behavior disappears and $v_{0},  v_{1}$ converge monotonically to 
the steady states reflecting the ranking $\epsilon_{0} < \epsilon_{1}$. 
\section{Summary and discussion}
\label{sec:sec4}
In this paper, we proposed 
a mathematical toy model, the so-called chiral Potts model to investigate the job-matching process in 
Japanese labor markets for university graduates and 
investigated the behavior analytically. 
We found several characteristic properties in the system. 
Let us summarize them below. 
For the case without ranking effect and market history, 
we observed that  the system undergoes fist order phase transition for $K \geq 3$ by 
changing the strength of cooperation $J (>0)$. 
When we take into account the ranking effect without market history, 
the ergodicity breaking region in $J$ appears. 
The market history affects on the dynamics of actual number of applicants to each company 
$v_{k}(t)$ to exhibit `zig-zag' behavior. 

We would like to stress that the situation and our modeling are applicable to 
the other type of resource allocation (utilization) such as the so-called {\it Kolkata Paise Restaurant (KPR) problem} \cite{KPR}.  
\subsection{Inverse problem of the Potts model}
However, from the view point of empirical science, 
in this model system,
the cross-correlations (the adjacent matrix) between students and companies are unknown and 
not yet specified.
Hence, we should estimate these elements by using appropriate empirical data sets. 
For instance,  if we obtain 
the `empirical correlation' $ \langle \delta_{\sigma_{i},\sigma_{j}} \rangle_{\rm emprical}$ 
from the data, we can determine $c_{ij}$ so as to satisfy 
the following relationship: 
\begin{eqnarray}
\langle \delta_{\sigma_{i},\sigma_{j}} \rangle & = &  
\frac{\partial}
{\partial c_{ij}}
\log \sum_{\bm{\sigma}}
{\exp}[-H(\bm{\sigma}: \{c_{ij}\})] \nonumber \\
\mbox{} & = &  
\frac{\sum_{\bm{\sigma}} \delta_{\sigma_{i},\sigma_{j}} 
{\exp}[-H(\bm{\sigma}: \{c_{ij}\})]}
{\sum_{\bm{\sigma}} 
{\exp}[-H(\bm{\sigma}: \{c_{ij}\})]}
= \langle \delta_{\sigma_{i},\sigma_{j}} \rangle_{\rm emprical}
\end{eqnarray}
where $\langle \delta_{\sigma_{i},\sigma_{j}} \rangle_{\rm emprical}$ might be evaluated empirically as 
a time-average by 
\begin{equation}
\langle \delta_{\sigma_{i},\sigma_{j}} \rangle_{\rm emprical} = 
(1/\tau)\sum_{t=t_{0}}^{\tau+t_{0}}
\delta_{\sigma_{i}^{(t)},\sigma_{j}^{(t)}}.
\end{equation}
We might also use the EM (Expectation and Maximization)-type algorithm \cite{Inoue2002} to infer the interactions. 
Those extensive studies in this directions (the `inverse Potts problem') 
including collecting the empirical data are now working in progress. 
\subsection{Learning of valuation basis of companies}
In this paper, 
we did not take into account the details of valuation process by companies so far. 
In our modeling, we assumed that they randomly select suitable students from the candidates up to their quota. 
This is because 
the valuation basis is unfortunately not opened for the public 
and it is somewhat `black box' for students. 
However, recently, 
several web sites \cite{Rikunabi,Mynabi} for supporting job hunting might collect 
a huge number of information 
about students as their `scores' of aptitude test. 

Hence, we might have a $N$-dimensional vector, 
each of whose component represents a score for a given question, 
for each student $l=1,\cdots,L$ as 
\begin{equation}
\bm{x}^{(l)}=
(x_{1}^{(l)},x_{2}^{(l)}, \cdots,x_{N}^{(l)})
\end{equation}
Then, we assume that 
each company $\mu=1,\cdots, K$ possesses 
their own valuation basis (weight) 
as a $N$-dimensional vector 
$\bm{a}_{\mu}=(a_{\mu 1},\cdots,a_{\mu N})$  and 
the score of student $l$ evaluated by the company $\mu=1,\cdots,K$ is given by 
\begin{equation}
y_{\mu}^{(l)}=
a_{\mu 1}x_{1}^{(l)}+a_{\mu 2}x_{2}^{(l)}+
\cdots + a_{\mu N}x_{N}^{(l)},\,\,\mu=1,\cdots,K. 
\end{equation}
It is naturally accepted that 
the company $\mu$ selects the students who are the $v_{\mu}^{*}$-top score candidates. 
Therefore, 
For a given threshold $\theta_{\mu}$, the decision by companies is given by 
\begin{equation}
\hat{y}_{\mu}^{(l)}=\Theta (y_{\mu}^{(l)}-\theta_{\mu}) =
\left\{
\begin{array}{cl}
1 & (\mbox{accept}) \\
0 & (\mbox{reject}) 
\end{array}
\right. 
\end{equation}
where $\Theta (\cdots)$ is a unit step function. 

Thus, for $L$ students and $K$ companies, 
the situation is determined by the following linear equation: 
\begin{equation}
\left(
\begin{array}{c}
y_{1}^{(l)} \\
\cdot \\
\cdot \\
\cdot \\
y_{M}^{(l)}
\end{array}
\right) = 
\left(
\begin{array}{ccccc}
a_{11} & \cdots  & \cdots & \cdots & a_{1N} \\
\cdots & \cdots & \cdots  & \cdots & \cdots \\
\cdots & \cdots & \cdots & \cdots & \cdots \\
\cdots & \cdots & \cdots & \cdots & \cdots \\
a_{M1} & \cdots & \cdots & \cdots & a_{MN}
\end{array}
\right)
\left(
\begin{array}{c}
x_{1}^{(l)} \\
\cdot \\
\cdot \\
\cdot \\
\cdot \\
\cdot \\
x_{N}^{(l)}
\end{array}
\right),\,\,\,l=1,\cdots,L
\end{equation}
namely, 
\begin{equation}
\bm{y}^{(l)}=
\bm{A}\bm{x}^{(l)},\,\,l=1,\cdots,L. 
\label{eq:matrix}
\end{equation}
When we have enough number of data sets 
$(\bm{y}^{(l)},\bm{x}^{(l)}),l=1,\cdots, L$, 
one might estimate the valuation base 
$\bm{A}$ by using suitable learning algorithm.  
When we notice that the above problem is described by `learning of a linear perceptron', 
one might introduce the following cost function: 
\begin{equation}
E = 
\frac{1}{2LM}
\sum_{l=1}^{L}
\sum_{\mu=1}^{M}
\delta_{s_{\mu}^{(l)},1}
\left\{
y_{\mu}^{(l)}-
\sum_{i=1}^{N}
a_{\mu i}x_{i}^{(l)}
\right\}^{2}
\end{equation}
where we defined 
$\delta_{a,b}$ as Kroneker's delta and  
\begin{equation}
s_{\mu}^{(l)}= 
\left\{
\begin{array}{cl}
1 & (\mbox{student $l$ sends an application letter to company $\mu$}) \\
0 & (\mbox{otherwise})
\end{array}
\right.
\end{equation}
Then,  we construct the learning equation as 
\begin{equation}
\frac{da_{\mu k}}{dt} = 
-\eta 
\frac{\partial E}{\partial a_{\mu k}}=
\frac{\eta}{LM}
\sum_{l=1}^{L}
\delta_{s_{\mu}^{(l)},1}
\left\{
y_{\mu}^{(l)}-
\sum_{i=1}^{N}
a_{\mu i}x_{i}^{(l)}
\right\}
x_{k}^{(l)}
\label{eq:Grad}
\end{equation}
for $\mu=1,\cdots, M, k=1,\cdots,N$. 

\begin{figure}[ht]
\begin{center}
\includegraphics[width=5.8cm]{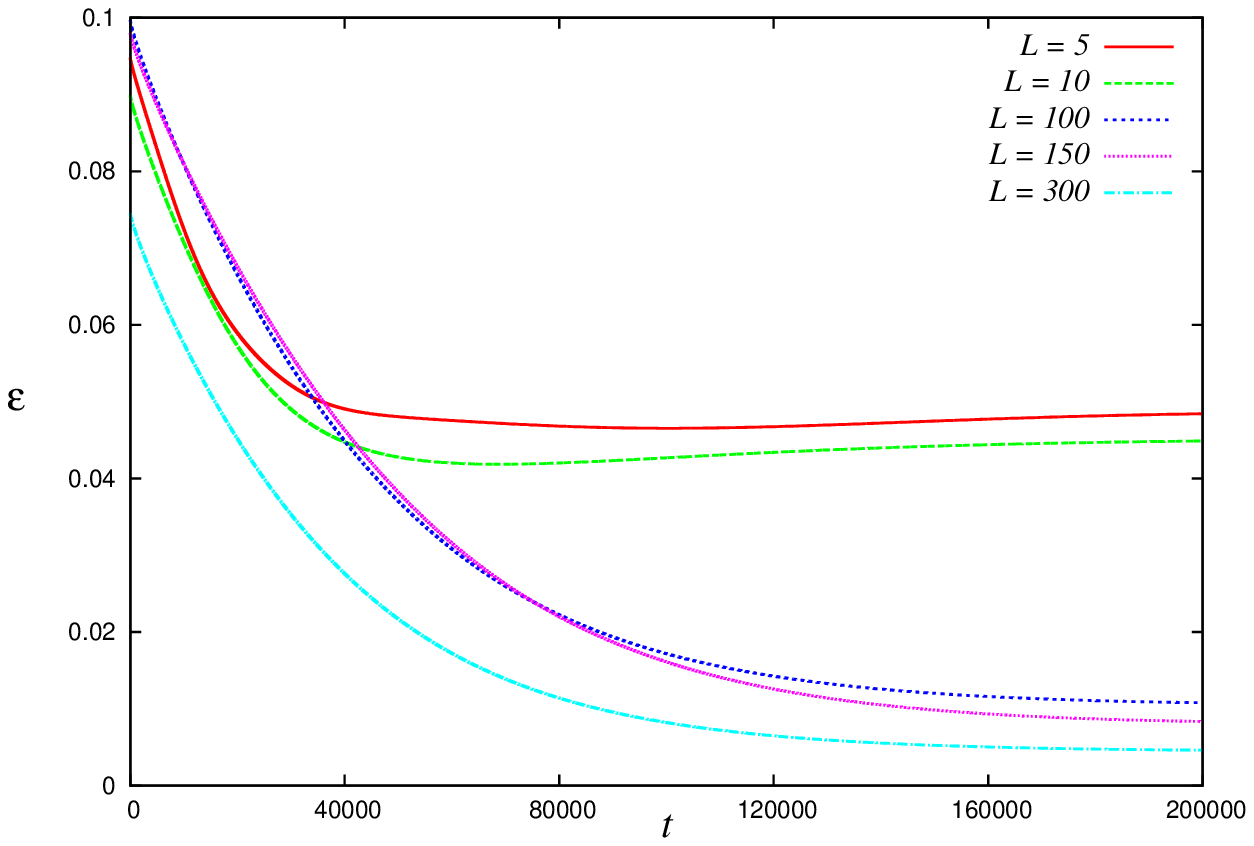} \hspace{-0.2cm}
\includegraphics[width=5.8cm]{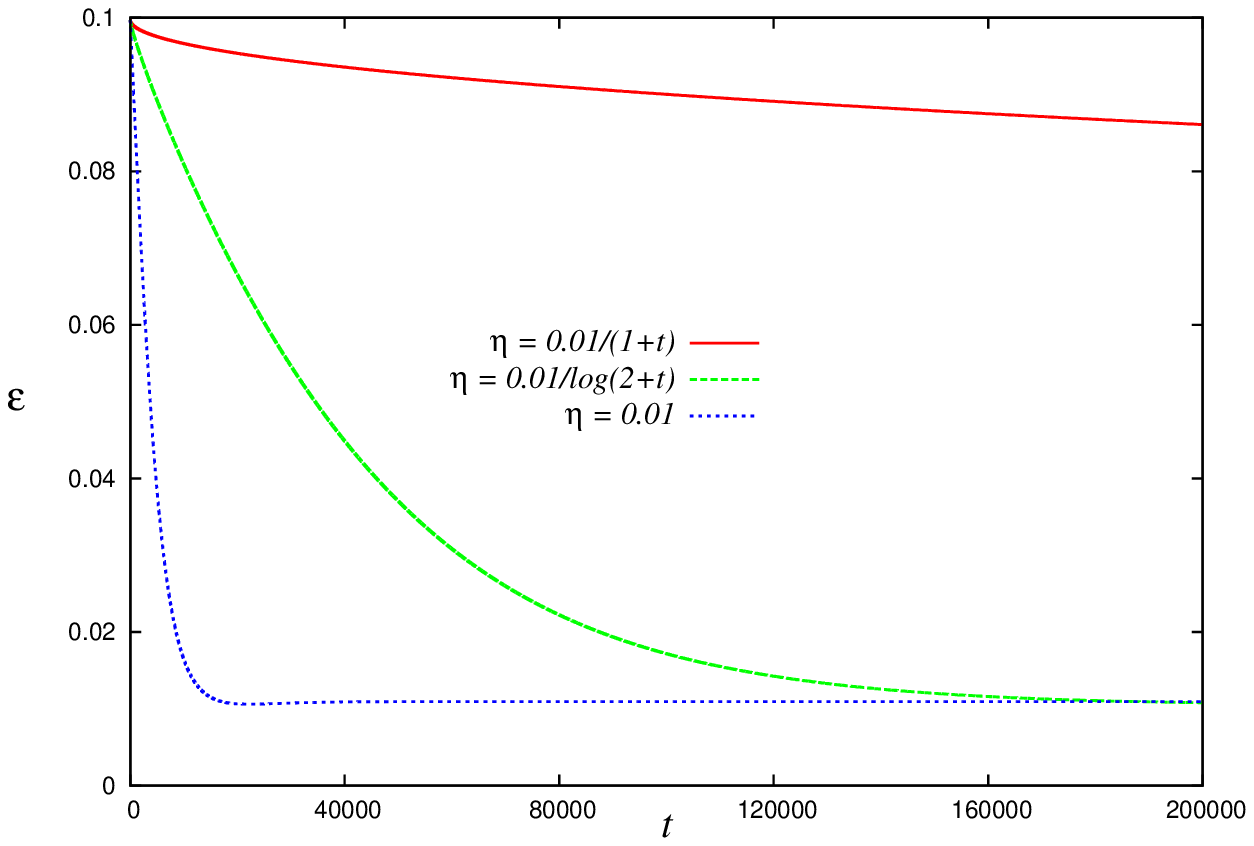}
\end{center}
\caption{\footnotesize 
Time-dependence of error 
$\epsilon (t)= 
(1/NM) 
\sum_{\mu=1}^{M}
\sum_{k=1}^{N}
(a_{\mu k}^{*}-a_{\mu k}(t))^{2}$ 
for the learning equation (\ref{eq:Grad}) 
using artificial data sets. 
We choose the learning rate as
$\eta=0.01/\log (2+t)$. 
$N=M=10$. 
}
\label{fig:fg_result1}
\end{figure}

We show an example of the learning dynamics 
through the error: 
\begin{equation}
\epsilon (t)= 
\frac{1}{NM}
\sum_{\mu=1}^{M}
\sum_{k=1}^{N}
(a_{\mu k}^{*}-a_{\mu k}(t))^{2}, 
\end{equation}
where $a_{\mu k}^{*}$ denotes a `true weight', 
for artificial data sets in Fig.  \ref{fig:fg_result1}. 
\mbox{}

Here we showed just an example of learning 
from artificial data sets for demonstration, 
however, 
it should be addressed as our future work 
to apply the learning algorithm to realistic situation using 
empirical data set collected from \cite{Rikunabi,Mynabi} 
or large-scale survey.
\mbox{}

Finally, it would be important for us to 
mention that it could be treated as 
`dictionary learning' \cite{Sakata} when 
the vector $\bm{x}^{(l)},l=1,\cdots, L$ is `sparse'  
in the context of {\it compressive sensing} \cite{Candes,Kabashima,Sompolinsky}. 
\subsubsection*{Acknowledgements}
This work was financially supported by Grant-in-Aid for Scientific Research (C) of
Japan Society for the Promotion of Science (JSPS) No. 2533027803 and 
Grant-in-Aid for 
Scientific Research on Innovative Area No. 2512001313. 
One of the authors (JI) thanks Hideaki Aoyama, Bikas K. Chakrabarti, Asim Ghosh, 
Siew Ann Cheong, Yoshi Fujiwara, 
Shigehiro Kato, 
Matteo Marsili and Subinay Dasgupta for fruitful discussion and useful comments. 
One of the authors (HC) was financially supported 
by Grant-in-Aid for the JSPS 
Research Fellowship for Young Scientists. 
We acknowledge organizers of 
{\it International Conference on 
Emerging Trends in Applied Mathematics}, 
in particular, 
Susmita Sarkar for 
warm hospitality during our stay in Kolkata and 
editing the conference proceedings. 


\end{document}